\documentclass[letter]{aa}

\usepackage{graphicx}
\usepackage{bm}
\usepackage{natbib}
\usepackage{mathtools}
\usepackage{amsmath}
\usepackage{multirow}
\usepackage[export]{adjustbox}
\usepackage{rotating}
\usepackage{tikz}
\usepackage{txfonts}
\usepackage{textcomp}
\begin{document} 

   \title{Undersampling effects on observed periods of coronal oscillations}
   
   \titlerunning{Undersampling effects on observed periods}

   \author{%
            {Daye Lim,}\inst{\ref{aff:CmPA}, \ref{aff:ROB}}
            {Tom Van Doorsselaere,}\inst{\ref{aff:CmPA}}
            {Valery M. Nakariakov}\inst{\ref{aff:Warwick}, \ref{aff:VIRAC}}
            {Dmitrii Y. Kolotkov}\inst{\ref{aff:Warwick}, \ref{aff:VIRAC}}
            {Yuhang Gao,}\inst{\ref{aff:Peking}, \ref{aff:CmPA}}
            \and {David Berghmans}\inst{\ref{aff:ROB}}
          }

   \institute{%
             \label{aff:CmPA}{Centre for mathematical Plasma Astrophysics, Department of Mathematics, KU Leuven, Celestijnenlaan 200B, 3001 Leuven, Belgium} \email{daye.lim@kuleuven.be}
            \and
            \label{aff:ROB}{Solar-Terrestrial Centre of Excellence – SIDC, Royal Observatory of Belgium, Ringlaan -3- Av. Circulaire, 1180 Brussels, Belgium}
            \and
            \label{aff:Warwick}{Centre for Fusion, Space and Astrophysics, Physics Department, University of Warwick, Coventry CV4 7AL, UK}
            \and
            \label{aff:VIRAC}{Engineering Research Institute \lq\lq Ventspils International Radio Astronomy Centre (VIRAC)\rq\rq, Ventspils University of Applied Sciences, Ventspils, LV-3601, Latvia}
            \and
            \label{aff:Peking}{School of Earth and Space Sciences, Peking University, Beijing, 100871, People\textquotesingle s Republic of China}
             }

   \authorrunning{Lim et al.}

  \abstract
   {Recent observations of decayless transverse oscillations have shown two branches in the relationship between periods and loop lengths. One is a linear relationship, interpreted as a standing mode. The other shows almost no correlation and has not yet been interpreted conclusively.}
   {We investigated the undersampling effect on observed periods of decayless oscillations.}
   {We considered oscillating coronal loops that closely follow the observed loop length distribution. Assuming that all oscillations are standing waves, we modeled a signal that represents decayless oscillations where the period is proportional to the loop length and the amplitude and phase are randomly drawn. A downsampled signal was generated from the original signal by considering different sample rates that mimic temporal cadences of telescopes, and periods for sampled signals were analysed using the fast Fourier transform.}
   {When the sampling cadence is getting closer to the actual oscillation period, a tendency for overestimating periods in short loops is enhanced. The relationship between loop lengths and periods of the sampled signals shows the two branches as in the observation.}
   {We find that long periods of decayless oscillations occurring in short loops could be the result of undersampling.}


\maketitle
\section{Introduction}
In the last decade, the Solar Dynamics Observatory/Atmospheric Imaging Assembly (SDO/AIA; \citealt{2012SoPh..275...17L}) and Solar Orbiter/Extreme Ultraviolet Imager (EUI; \citealt{2020A&A...642A...8R}), with high spatial resolution and unprecedented high temporal cadence, have provided plenty of observations of decayless transverse oscillations of solar coronal loops. These oscillations have relatively small displacement amplitudes, comparable to or smaller than the minor radius of the oscillating loop, without significant damping and are described well with a harmonic function \citep{2012ApJ...751L..27W, 2012ApJ...759..144T, 2021SSRv..217...73N, 2022MNRAS.513.1834Z, 2022MNRAS.516.5989Z, 2022A&A...666L...2M, 2023NatCo..14.5298Z, 2023NatSR..1312963Z, 2023ApJ...946...36P, 2023ApJ...944....8L, 2024A&A...685A..36S}. The decayless oscillations are considered a common feature in the solar corona by being excited without intermittent solar energetic events \citep{2015A&A...583A.136A, 2022ApJ...930...55G, 2024arXiv240606368L}.  

In the case of decayless oscillations occurring in longer loops with an average length of about 200 Mm, the statistical result showed that the observed periods ranging from around 1.5 min to 10 min have a clear linear relationship with the loop lengths \citep{2013A&A...560A.107A, 2013A&A...552A..57N, 2015A&A...583A.136A, 2022A&A...666L...2M, 2022MNRAS.513.1834Z, 2023ApJ...944....8L}. This is in agreement with a characteristic of a standing mode. The interpretation as a standing mode was established as additional observations, e.g., consistent oscillation phase \citep{2015A&A...583A.136A} and multiple harmonics \citep{2018ApJ...854L...5D}, were discovered. 

The decayless oscillations were also observed in shorter loops with an average length of about 20 Mm, and the lengths of the oscillating loops showed no correlation with periods ranging between about 10~s and 5~min \citep{2022ApJ...930...55G, 2024A&A...685A..36S}. Attempts were made to find phase relations in oscillations along the loop axis for interpreting these oscillations \citep{2023ApJ...946...36P, 2024A&A...685A..36S}, however, the short lengths prevented the authors from concluding that they were either standing or propagating from an observational perspective. Theoretical and numerical studies have begun to pay attention to understanding these oscillations with periods much longer than expected from their loop lengths. \citet{2023ApJ...955...73G} found that decayless oscillations with a long period of about 5 min in a short coronal loop with a length of 30 Mm could be directly driven by the photospheric p-mode driver. \citet{2024MNRAS.527.5741L} suggested that decayless oscillations having longer periods in shorter loops could be the manifestation of slow mode oscillations driven by the p-modes.

In addition to the possible physical models that can explain long-period decayless oscillations in short loops, in this letter, we show that they can also be caused by undersampling. In signal processing, undersampling occurs when a sample rate to receive signal data is insufficient to reconstruct a real signal. This causes a false spectral peak at a frequency lower than the actual frequency. Even though we now image the Sun relatively continuously at a faster cadence (e.g., from 12~s by AIA down to 2~s by EUI) than ever before, this could still be insufficient to detect the short-period oscillations that occur in small-scale loops.

\section{Results}\label{sec:result}
\begin{figure}
  \resizebox{\hsize}{!}{\includegraphics{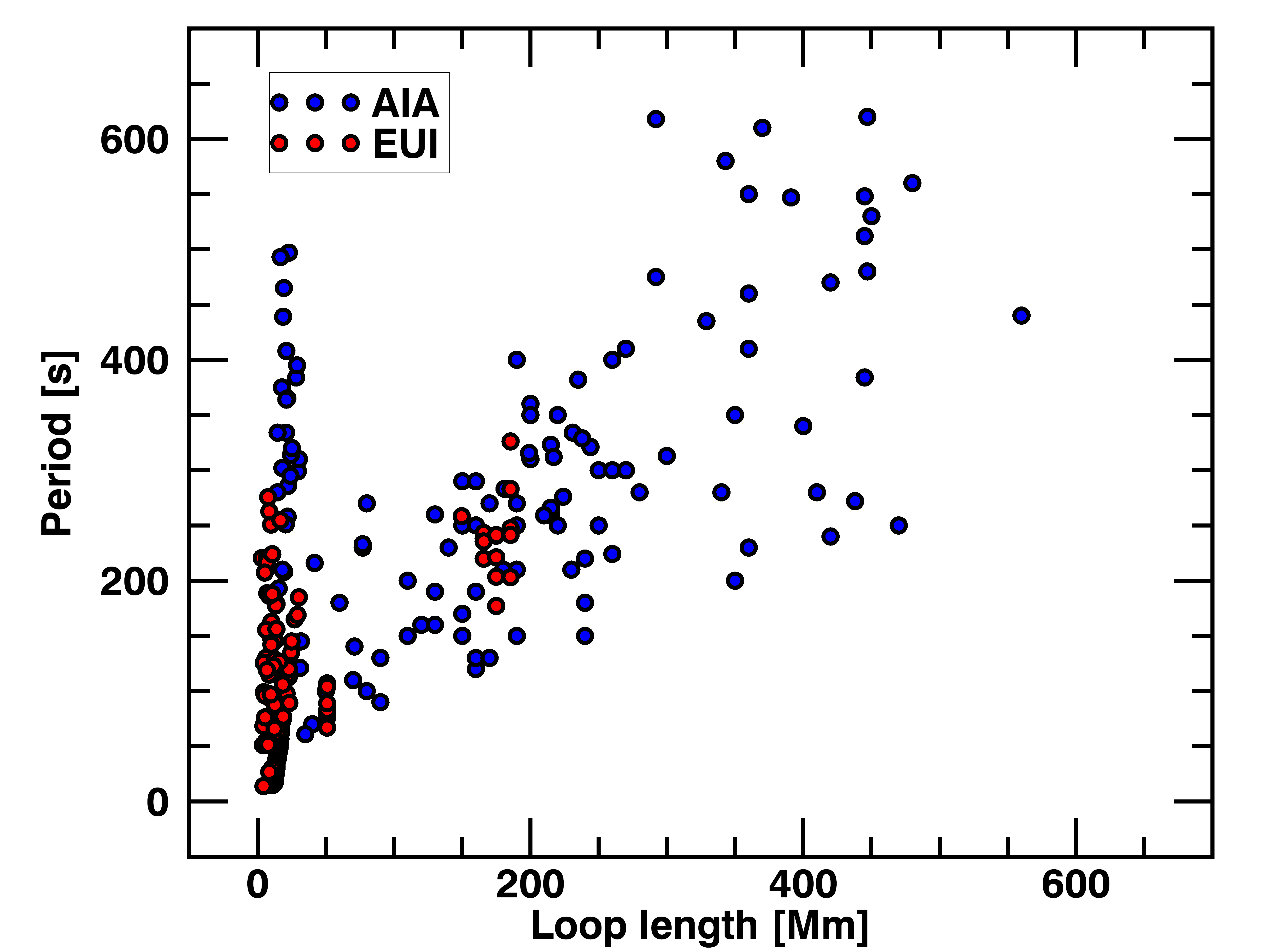}}
  \caption{Scatter plot of periods and loop lengths of decayless transverse oscillations of coronal loops observed by SDO/AIA (blue) and Solar Orbiter/EUI (red).}
  \label{fig:len_per_obs}
\end{figure}

In order to model oscillating coronal loops and their oscillations, we consider observed parameters in decayless transverse oscillations in the literature \citep{2012ApJ...751L..27W, 2013A&A...552A..57N, 2013A&A...560A.107A, 2015A&A...583A.136A, 2018ApJ...854L...5D, 2019ApJ...884L..40A, 2022ApJ...930...55G, 2022MNRAS.513.1834Z, 2022A&A...666L...2M, 2022MNRAS.516.5989Z, 2023ApJ...944....8L, 2023ApJ...946...36P, 2023NatSR..1312963Z, 2024A&A...685A..36S}. These oscillations were observed by AIA with the cadence of 12-s and EUI with the cadence of 2, 3, and 5~s. As already mentioned above, the relationships between lengths and periods showed two branches (Figure \ref{fig:len_per_obs}). One follows a linear function with a slope of approximately 0.8 $\text{s}\,\text{Mm}^{-1}$, and the other is distributed almost vertically between loop lengths of 0 and 50 Mm.

\begin{figure*}
  \resizebox{\hsize}{!}{\includegraphics{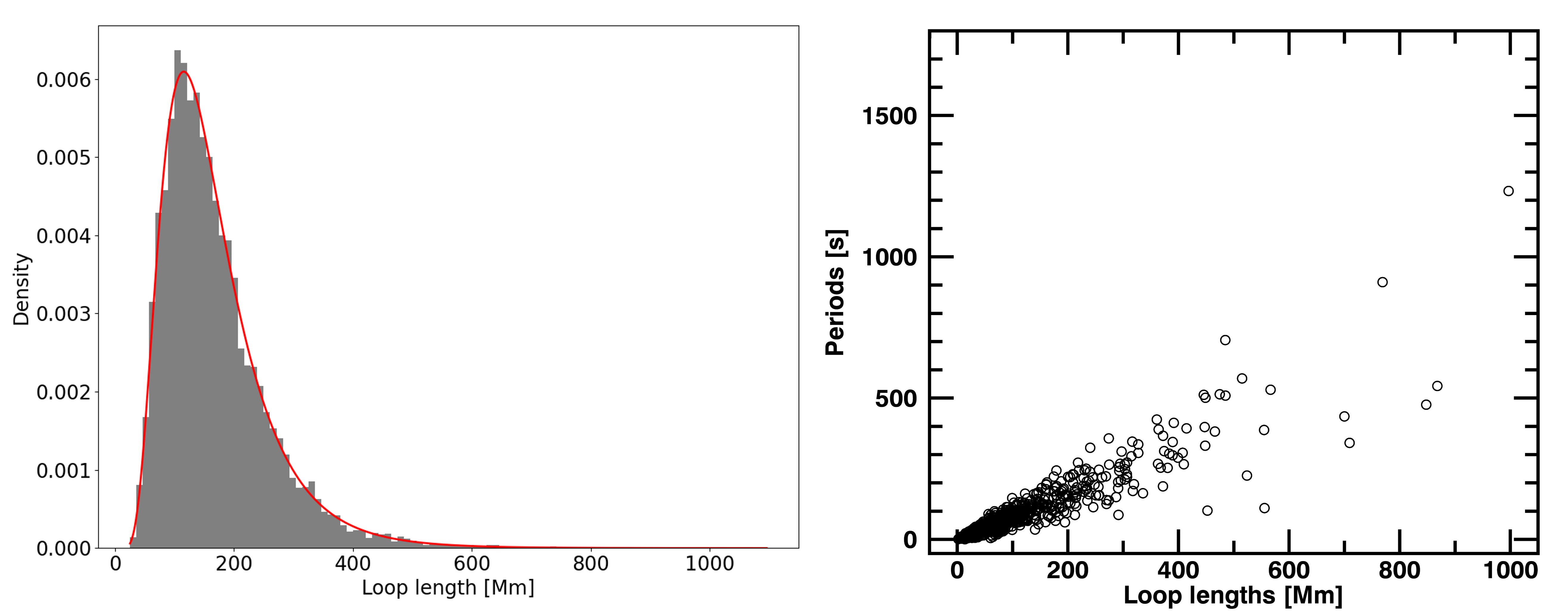}}
  \caption{Loop length distribution (left panel) and the scatter plot between the loop lengths and periods derived from them (right panel).}
  \label{fig:dist}
\end{figure*}

Using a Monte Carlo approach, we consider 1000 lengths ($L$) randomly drawn from a log-normal distribution $\text{ln}N(5, 0.5^2)$ which gives the arithmetic mean loop length of about 170 Mm, and the standard deviation of around 90 Mm (see the left panel of Figure~\ref{fig:dist}). The choices of the mean, standard deviation, and distribution type are driven by the observed loop lengths shown in Figure~\ref{fig:len_per_obs}. We assume that all decayless oscillations occurring in the loops are in standing mode. Hence, the period should increase linearly with the loop length. To simulate the linear branch in the observed relationship between lengths and periods, we consider the periods ($P$) as follows
\begin{equation}
    P=0.8LN(1, 0.3^2),
\end{equation}
where $N(1, 0.3^2)$ is a normal distribution with the mean of 1 $\text{s}\,\text{Mm}^{-1}$ and the standard deviation of 0.3 $\text{s}\,\text{Mm}^{-1}$, and all the parameters were justified by observations. The right panel of Figure \ref{fig:dist} shows that the simulated period and loop length well mimic the linear relationship between the observed period and loop length. Using these periods, we model 1000 signals ($S$) for decayless oscillations described by a sine function,
\begin{equation}
    S_{i}(t)=A_{i}\text{sin}(2\pi t/P_{i}+\phi_{i}),
\end{equation}
where $i$ is an index for each signal, $x$ and $t$ are a position and time, $A$ is the displacement amplitude randomly drawn ranging between 10 and 400~km according to observed parameters, and $\phi$ is a temporal phase randomly drawn. Each signal has 3600 data points, and we assume that the time difference between two data points is equally 1-s to help us understand by converting it into a period we are familiar with.

\begin{figure*}
  \resizebox{\hsize}{!}{\includegraphics{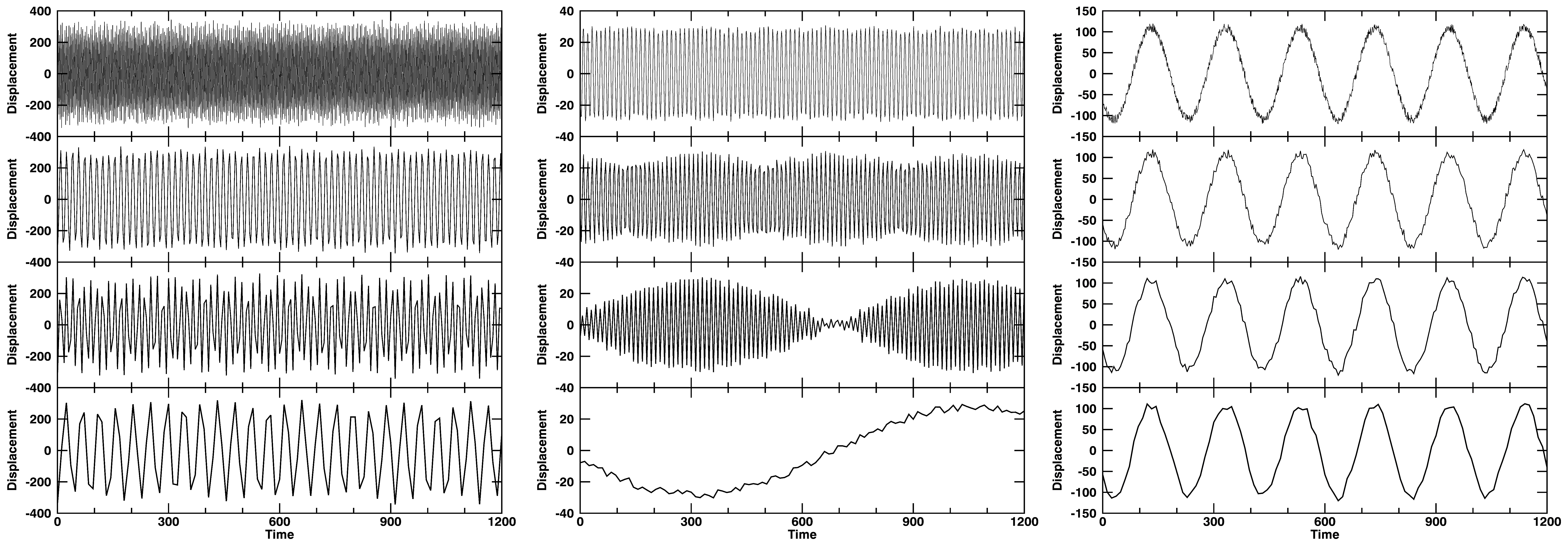}}
  \caption{Examples of original signals (top), sampled signal at 3-s (second row), 6-s (third row), and 12-s (fourth row). Left panels: true periods and detected periods from sampled signals from top to bottom are 1, 16, 16, and 45 s. Middle panels: 12, 12, 12, and 1455 s. Right panels: 200, 200, 200, and 200 s.}
  \label{fig:signal}
\end{figure*}

\begin{figure*}
  \resizebox{\hsize}{!}{\includegraphics{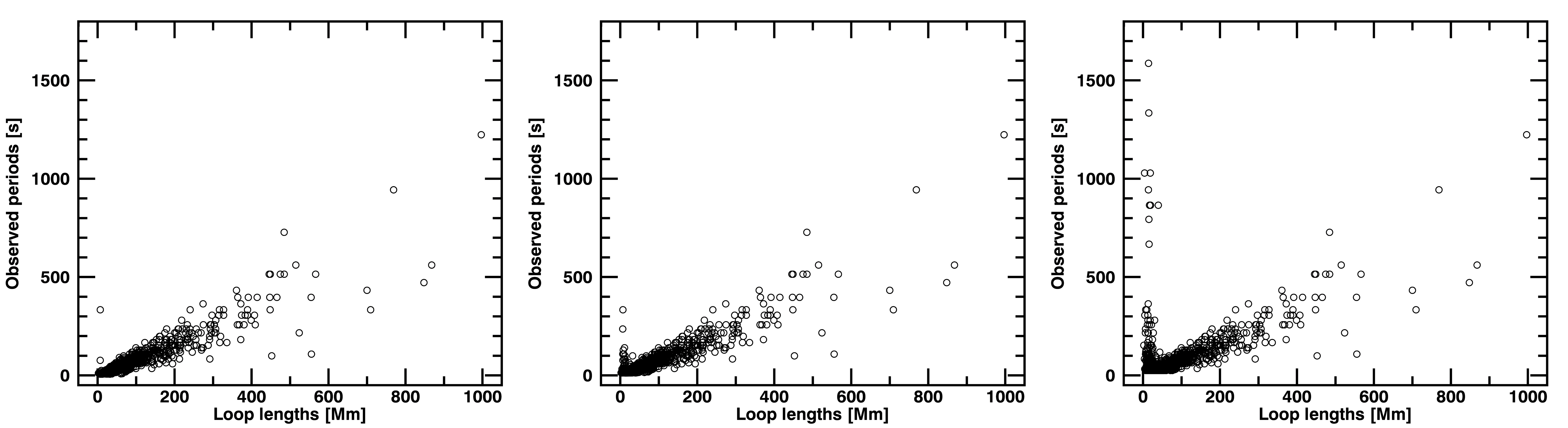}}
  \caption{Scatter plots between loop lengths and detected periods from sampled signals at 3 data points (left panel), 6 data points (middle panel), and 12 data points (right panel).}
  \label{fig:len_per_model}
\end{figure*}

In order to investigate the effect of undersampling on detected periods, we sample each signal at different rates, i.e., we select signal data at 3, 6, and 12~s. The different sampling rates mimic different temporal cadences of imaging observations. Figure \ref{fig:signal} presents examples of original signals and sampled signals. We can see that the signals show different apparent periodicities depending on sampling cadences. We investigate a periodicity for sampled signals using a fast Fourier transform (FFT) with a 95\% significance level. The relationship between detected periods for each sampled case and loop lengths is shown in Figure~\ref{fig:len_per_model}. For short loop lengths between about 0 and 50 Mm, longer periods are found depending on the sampling cadence compared to the original relationship (Figure~\ref{fig:dist}), and this effect is further enhanced as the sampling cadence increases, i.e., when the cadence time approaches the oscillation period. The relationship between the loop length and period of the signals sampled at 12-s correctly reconstructs the two branches that appear in the observed relationship shown in Figure~\ref{fig:len_per_obs}. It would be worth mentioning here that the detected periods are not affected by the time difference between the two data points, i.e., the number of data points for one period, and the total duration of signals.

\section{Implications of the proposed interpretation}\label{sec:conclusion}
Our results indicate that when interpreting long-period decayless oscillations in short loops, there is a possibility that they are falsely detected as long-periods due to undersampling, when in fact they are standing modes with short periods. This effect should be taken into consideration. Considering Nyquist's theorem stating that a periodic signal must be sampled at more than twice the highest frequency of the signal, the shortest period that we can detect without undersampling is 24~s for AIA and 6~s for EUI (with a 3-s cadence). In other words, periods shorter than 24~s (6~s) may be incorrectly detected as longer periods in AIA (EUI) observations. 

The oscillations with a period of longer than about 150~seconds in loop length between 0 and 50~Mm (belonging to the branch that deviates from the linear relationship) were detected with both AIA \citep{2022ApJ...930...55G} and EUI \citep{2023ApJ...944....8L, 2024A&A...685A..36S}. 
The average length of the loops, which have the oscillations corresponding to this branch observed by EUI, is approximately 10 Mm. If a standing mode following the linear relationship shown in Figure~\ref{fig:len_per_obs} occurs in this loop, its period would be about 8~s, which is detectable with the 3-second cadence of EUI according to the Nyquist theory. However, the possibility of an undersampling effect in EUI cannot be ruled out. For example, \citet{2024A&A...685A..36S} detected an oscillation with a period of approximately 220 seconds in a loop about 3 Mm in length. This oscillation was observed with a 3-second cadence in HRIEUV. It is possible that the actual period could be closer to 5 s (suggesting a phase speed of 1200~km $\text{s}^{-1}$ under the assumption of a standing mode), which would be undetectable by HRIEUV due to the Nyquist frequency. Therefore, the long-period oscillations in short loops detected with EUI could be driven by physical mechanisms and may also be influenced by the undersampling effect.
In the case of AIA, the oscillations following the branch that is not the linear relationship occurred in loops with an average length of around 20 Mm. A possible period of a standing mode in this loop would be approximately 16~s, which is not possible to detect with the 12-s cadence of AIA given the Nyquist frequency. Therefore, the long-period oscillations in short loops observed by AIA could be the effect of undersampling and their periods could be actually shorter than the period that can be detected by the Nyquist theory, i.e., shorter than 24-s.

This interpretation has implications for the seismological estimation discussed in \citet{2022ApJ...930...55G}. They reported 30 decayless oscillations with a period of longer than 100~s in small-scale coronal loops observed with AIA. When interpreting these oscillations as the fundamental kink mode, the derived phase speed has an average value of about 200~km $\text{s}^{-1}$, and the average magnetic field strength measured by seismological inference was about 3 G, which were all smaller than typical coronal values. Considering that the period that can be detected by AIA without undersampling effects is 24~s or longer, it is possible that these oscillations were incorrectly detected because their periods were shorter than 24~s. If we assume that their true periods are approximately 12~s, the inferred average magnetic field strength is about 50 G, which is closer to the coronal magnetic field in active regions. 

Moreover, decayless oscillations have received much attention in terms of the energy content that they can contribute to coronal heating \citep{2022ApJ...930...55G, 2023ApJ...946...36P, 2023ApJ...944....8L, 2024A&A...685A..36S}. The energy flux ($F$) of standing transverse oscillation is proportional to
\begin{equation}
    F\propto \left(\frac{2\pi A}{P}\right)^{2}\left(\frac{2L}{P}\right).
\end{equation}
An overestimated period leads to an underestimated energy flux. \citet{2023ApJ...952L..15L} performed a meta-analysis and estimated the average energy flux of the oscillations detected by \citet{2022ApJ...930...55G}, to be about 0.2~$\text{W}\,\text{m}^{-2}$. However, assuming that they are all a standing mode with a short period of about 12-s as assumed above, the average energy flux is estimated to be about 800~$\text{W}\,\text{m}^{-2}$. It is important to note that this value represents an upper limit, and the actual dissipated energy is likely to be lower. Utilizing the analytical model of dissipation rates presented by \citet{2020ApJ...897L..13H}, we calculated the dissipation rates for 31 oscillations from Gao et al. (2022), using the corrected periods. The dissipation rates ranged from approximately 1\% to 60\%, with an average dissipated energy flux of around 2~$\text{W}\,\text{m}^{-2}$.

We do not rule out at all the possibility that long periods occurring in short loops could be driven by physical mechanisms. However, we would like to emphasize that observed oscillation parameters, such as periods, are used in seismological tools and energy content estimation, requiring a precise understanding of the observed oscillations, including the physical mechanisms as well as uncertainties arising from the limitations of the instrument specifications. Solar Orbiter/EUI allows us to investigate coronal structures at smaller and smaller scales, which requires the highest cadence possible to avoid the effects of undersampling. We could also expect this from future missions targeting high resolution and high temporal cadence, such as the Multi-slit Solar Explorer (down to 0.5~s; \citealt{2022ApJ...926...52D}) and Solar-C EUV High-Throughput Spectroscopic Telescope (maximum at 1~s; \citealt{2019SPIE11118E..07S}). Observational support for the arguments presented in this paper could be achieved by comparing the detected periods of oscillations observed simultaneously with high-cadence EUI and AIA. We will pursue this approach in the future when more suitable datasets become available.

\begin{acknowledgements}
      Solar Orbiter is a space mission of international collaboration between ESA and NASA, operated by ESA. The EUI instrument was built by CSL, IAS, MPS, MSSL/UCL, PMOD/WRC, ROB, LCF/IO with funding from the Belgian Federal Science Policy Office (BELSPO/PRODEX PEA C4000134088); the Centre National d’Etudes Spatiales (CNES); the UK Space Agency (UKSA); the Bundesministerium für Wirtschaft und Energie (BMWi) through the Deutsches Zentrum für Luft- und Raumfahrt (DLR); and the Swiss Space Office (SSO). DL was supported by a Senior Research Project (G088021N) of the FWO Vlaanderen. TVD was supported by the C1 grant TRACEspace of Internal Funds KU Leuven and a Senior Research Project (G088021N) of the FWO Vlaanderen. Furthermore, TVD received financial support from the Flemish Government under the long-term structural Methusalem funding program, project SOUL: Stellar evolution in full glory, grant METH/24/012 at KU Leuven. The research that led to these results was subsidised by the Belgian Federal Science Policy Office through the contract B2/223/P1/CLOSE-UP. The paper is also part of the DynaSun project and has thus received funding under the Horizon Europe programme of the European Union under grant agreement (no. 101131534). Views and opinions expressed are however those of the author(s) only and do not necessarily reflect those of the European Union and therefore the European Union cannot be held responsible for them.
      We also acknowledge funding from the STFC consolidated grant ST/X000915/1 (DYK) and Latvian Council of Science Project No. lzp2022/1-0017 (DYK and VMN).
      Y.G. was supported by China Scholarship Council under file No. 202206010018.
\end{acknowledgements}

\bibliographystyle{aa} 
\bibliography{Lim_bib} 

\end{document}